\documentstyle[12pt,epsf,epsfig,amsmath]{article}
\newfont{\thiplo}{msbm10 scaled\magstep 2}
\newfont{\gothic}{eufb10 scaled\magstep 2}
\newfont{\unc}{eurb10} 
\newskip\humongous \humongous=0pt plus 1000pt minus 1000pt
\def\caja{\mathsurround=0pt}
\def\eqalign#1{\,\vcenter{\openup1\jot \caja
        \ialign{\strut \hfil$\displaystyle{##}$&$
        \displaystyle{{}##}$\hfil\crcr#1\crcr}}\,}
\newif\ifdtup

\def\eqright #1\cr{\noalign{\hfill$\displaystyle{{}#1}$}}
\def\eqleft #1\cr{\noalign{\noindent$\displaystyle{{}#1}$\hfill}}
\def\oldreffmt#1{\rlap{[#1]} \hbox to 2\parindent{}}

\def\figfmt#1{\rlap{Figure {#1}} \hbox to 1in{}}

\def\sectioneq{\def\theequation{\thesection.\arabic{equation}}{\let
\holdsection=\section\def\section{\setcounter{equation}{0}\holdsection}}}%

\newcounter{holdequation}

\def\begineq #1\endeq{$$ \refstepcounter{equation}\eqalign{#1}\eqno
	(\theequation) $$}
\def\contlimit{\,{\hbox{$\longrightarrow$}\kern-1.8em\lower1ex
\hbox{${\scriptstyle (a\rightarrow0)}$}}\,}
\def\centeron#1#2{{\setbox0=\hbox{#1}\setbox1=\hbox{#2}\ifdim
\wd1>\wd0\kern.5\wd1\kern-.5\wd0\fi
\copy0\kern-.5\wd0\kern-.5\wd1\copy1\ifdim\wd0>\wd1
\kern.5\wd0\kern-.5\wd1\fi}}
\def\centerover#1#2{\centeron{#1}{\setbox0=\hbox{#1}\setbox
1=\hbox{#2}\raise\ht0\hbox{\raise\dp1\hbox{\copy1}}}}
\def\centerunder#1#2{\centeron{#1}{\setbox0=\hbox{#1}\setbox
1=\hbox{#2}\lower\dp0\hbox{\lower\ht1\hbox{\copy1}}}}
\def\lsim{\;\centeron{\raise.35ex\hbox{$<$}}{\lower.65ex\hbox
{$\sim$}}\;}
\def\gsim{\;\centeron{\raise.35ex\hbox{$>$}}{\lower.65ex\hbox
{$\sim$}}\;}
\def\super#1{\ifmmode \hbox{\textsuper{#1}}\else\textsuper{#1}\fi}
\def\textsuper#1{\newcount\holdspacefactor\holdspacefactor=\spacefactor
$^{#1}$\spacefactor=\holdspacefactor}
\def\getcite#1,{\advance\citenumber by1
\def\getcitearg{#1}\def\lastarg{@}
\ifnum\citenumber=1
\ref{#1}\let\next=\getcite\else\ifx\getcitearg\lastarg\let\next=\relax
\else ,\ref{#1}\let\next=\getcite\fi\fi\next}
\def\pom{{\rm P\kern -0.53em\llap I\,}}
\def\spom{{\rm P\kern -0.36em\llap \small I\,}}
\def\sspom{{\rm P\kern -0.33em\llap \footnotesize I\,}}

\relax
\def\contlimit{\,{\hbox{$\longrightarrow$}\kern-1.8em\lower1ex
\hbox{${\scriptstyle (a\rightarrow0)}$}}\,}
\def\upon #1/#2 {{\textstyle{#1\over #2}}}
\relax
\renewcommand{\thefootnote}{\fnsymbol{footnote}}

\sectioneq
\def\subhead#1{\bigskip\vbox{\noindent\bf #1}\nobreak\par}

\def\til#1{\centeron{\hbox{$#1$}}{\lower 2ex\hbox{$\char'176$}}}
\def\tild#1{\centeron{\hbox{$\,#1$}}{\lower 2.5ex\hbox{$\char'176$}}}
\def\sumtil{\centeron{\hbox{$\displaystyle\sum$}}{\lower
-1.5ex\hbox{$\widetilde{\phantom{xx}}$}}}

\begin{document} 

\begin{titlepage} 

\rightline{\vbox{\halign{&#\hfil\cr
&ANL-HEP-CP-04-83\cr
&\today\cr}}} 
\vspace{0.25in} 

\begin{center} 
  
{\large\bf The Sextet Higgs Mechanism and the Pomeron}\footnote{Presented at the 
International Symposium on Multiparticle Dynamics, Sonoma, CA, 
July 2004.}\footnote{Work 
supported by the U.S.
Department of Energy under Contract
W-31-109-ENG-38} 

\medskip

Alan. R. White\footnote{arw@hep.anl.gov }

\vskip 0.6cm

\centerline{Argonne National Laboratory}
\centerline{9700 South Cass, Il 60439, USA.}
\vspace{0.5cm}

\end{center}

\begin{abstract} 

If electroweak symmetry breaking is a consequence of
color sextet quark chiral symmetry breaking,
dramatic, large cross-section, effects are to be expected at the LHC 
- with the pomeron playing a prominent role. The symmetry breaking is tied 
to a special solution of QCD which can be constructed, at high-energy,
via the chiral anomaly and  reggeon diagrams. 
There is confinement and chiral symmetry
breaking, but physical states contain both quarks and  
a universal, anomalous, wee gluon component.
A variety of Cosmic Ray effects could be supporting evidence,  
including the knee in the spectrum
and the ultra-high energy events. 
The sextet neutron should be stable and is a natural 
dark matter candidate. A large  
$E_T$ jet excess at Fermilab, and large $x$ and $Q^2$ events at HERA,
would be supporting accelerator evidence. Further evidence, including
diffractive-related vector boson pair production and 
top quark related phenomena, could be seen at Fermilab 
as data is accumulated.

\end{abstract} 

\renewcommand{\thefootnote}{\arabic{footnote}} \end{titlepage}

\subhead{ 1. Introduction.}

By pursuing a consistent description of high-energy QCD via the
critical Pomeron we have been led$^1$ 
to the radical proposition that there should be a second (higher color) quark 
sector of the theory, with electroweak scale masses and even stronger 
interactions. In addition to providing a unitary high-energy
solution of QCD, the existence of this sector would naturally link, 
and potentially 
solve, some of the most prominent problems of current-day astro-particle 
physics that, at first sight, are unrelated to QCD. Included are, the origin 
of electroweak symmetry breaking, the presence and dominance of dark matter,
the knee in the cosmic ray spectrum and, perhaps, the origin and nature of 
ultra-high energy cosmic rays. In accelerator physics
the first experimental evidence should be subtle, electroweak scale, 
deviations from standard QCD, that may have already been seen.
At the LHC, there should be a multitude of large cross-section effects.
Double pomeron exchange could produce 
the most immediately observable (definitive) effect and may be spectacular!

\subhead{ 2. The Sextet Quark Higgs Mechanism.} 

A color sextet quark doublet (plus heavy leptons - to most easily
avoid an SU(2)xU(1) anomaly)
produces electroweak symmetry breaking in a manner that, 
at first sight, is simply a special version of technicolor, with QCD as the 
technicolor gauge group. 
Breaking of the sextet chiral symmetry
gives a triplet $\Pi^{\pm}$, $\Pi^0$
of sextet pions and a ``Higgs'' (the $\eta_6$)
and the $W^{\pm}$ and $Z^0$ acquire masses by ``eating'' the $\Pi$'s.
But, economically and very beautifully,
no new interaction is needed (beyond SU(3)xSU(2)xU(1) gauge interactions).
Instead, the electroweak scale is a second (higher color) QCD scale 
and electroweak symmetry breaking is intricately connected
with QCD dynamics, with the direct 
QCD production of $\Pi$'s giving a range of new phenomena.

\subhead{ 3. QCD$_S$} 

$QCD_S$ (with six color triplet and two color sextet quarks)
has several special properties, compared to conventional QCD.
An infra-red fixed point and the possibility 
to add an asymptotically-free scalar field allows a unitary 
high-energy S-Matrix to be constructed ``diagrammatically'' 
by starting with SU(3) color broken to SU(2) ($CSQCD_S$). As we describe below,
there is confinement and chiral symmetry
breaking but (infinite momentum) physical states contain both quarks and  
a crucial (universal) ``anomalous wee gluon'' component.
The pomeron is (approximately) a regge pole and the ``Critical Pomeron'' 
describes asymptotic high-energy cross-sections. 

A priori, the $\eta_6$ appears to be 
a light axion of the kind ruled out experimentally.
Fortunately, in $QCD_S$ the wee gluon 
content of the states breaks the associated U(1)
symmetry and there is no light axion in the ``anomaly S-Matrix''. Instead, 
the $\eta_6$ should have an electroweak scale mass
and could be responsible for top quark production.
For related reasons, there should be no glueball asymptotic states,
no BFKL pomeron, and no odderon. 
The only new baryons will be the ``sextet proton'' ($P_6$)
and the ``sextet neutron'' ($N_6$). The $N_6$ should be stable and 
is a strong candidate for dark matter and also for the highest energy cosmic rays.

\subhead{4. High-Energy QCD$_S$ Via Color Superconductivity.}

That the regge behavior of $QCD_S$
should be constructed via the ($k_{\perp}$ cut-off) reggeon diagrams of 
$CSQCD_S$ was first 
motivated by a match with supercritical pomeron diagrams.
We now understand that this allows the chiral 
anomaly structure produced by longitudinal, massive, gluons to play a crucial
role and resolves the (Gribov) ambiguity
associated with infinite momentum longitudinal wee gluons in unbroken QCD.

We take all quarks  massless\footnote{We will not discuss 
whether effective quark masses can be generated by the sextet
Higgs mechanism. Significant shifting of the Dirac sea, perhaps 
as envisaged by Gribov, has to be involved.}, so that
the pinching of quark and antiquark
propagator poles produces zero momentum chirality transitions, 
via the anomaly, in the regge limit ``effective vertex'' triangle diagrams
that appear in $CSQCD_S$. With a $k_{\perp}$ cut-off,
the chirality transitions produce wee gluon divergences at  $k_{\perp}=0$.
Because of the infra-red fixed-point and regge exponentiation, these divergences  
remove all colored wee gluon states. An overall divergence
that does not exponentiate, produced by color zero anomalous
(charge parity $\neq$ signature) wee gluons, selects the physical
amplitudes. The divergence is factored out as a $k_{\perp}=0$
``condensate''and the resulting physical states
are ``anomaly pole'' Goldstone bosons (``pions'').

The simplest $\pi - \pi$ scattering diagrams have the form shown in Fig.~1, 
where the dynamical contribution of the chirality transitions is also illustrated. 
In effect, an anomaly pole ``pion''
is created by the product of a physical quark field and a zero momentum
``unphysical'' antiquark field in which the Dirac sea is shifted. 
Via a chirality  transition, the antiquark becomes 
physical in the presence of a compensating (condensate) wee gluon field.
(There is no pure  $~q\bar{q}$ component 
in the infinite momentum pion ``wave function'' 
- the $q\bar{q}~$ pair has  vector-like spin!) In the scattering a 
further chirality transition re-aligns the initial state
wee gluon field for absorption by the final state pion. 
The condensate can be viewed as 
a shift of the Dirac sea that produces an 
S-Matrix in which
confinement of SU(2) color and chiral symmetry breaking completely determine
the spectrum of states.

\noindent \parbox{3.7in}{
$~~$ \epsfxsize=1.3in
\epsffile{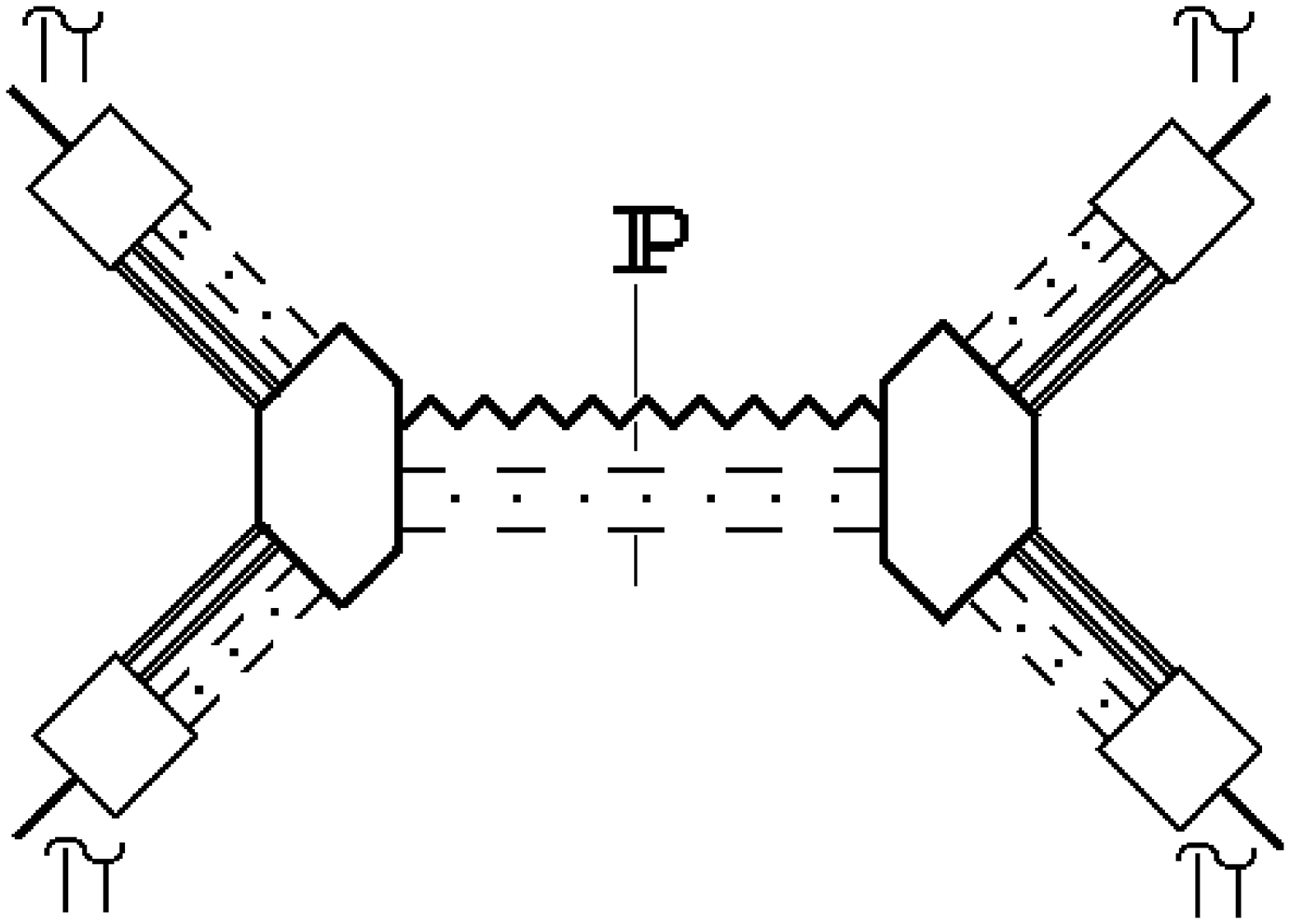}
\hspace{0.1in}
\epsfxsize=2.1in
\epsffile{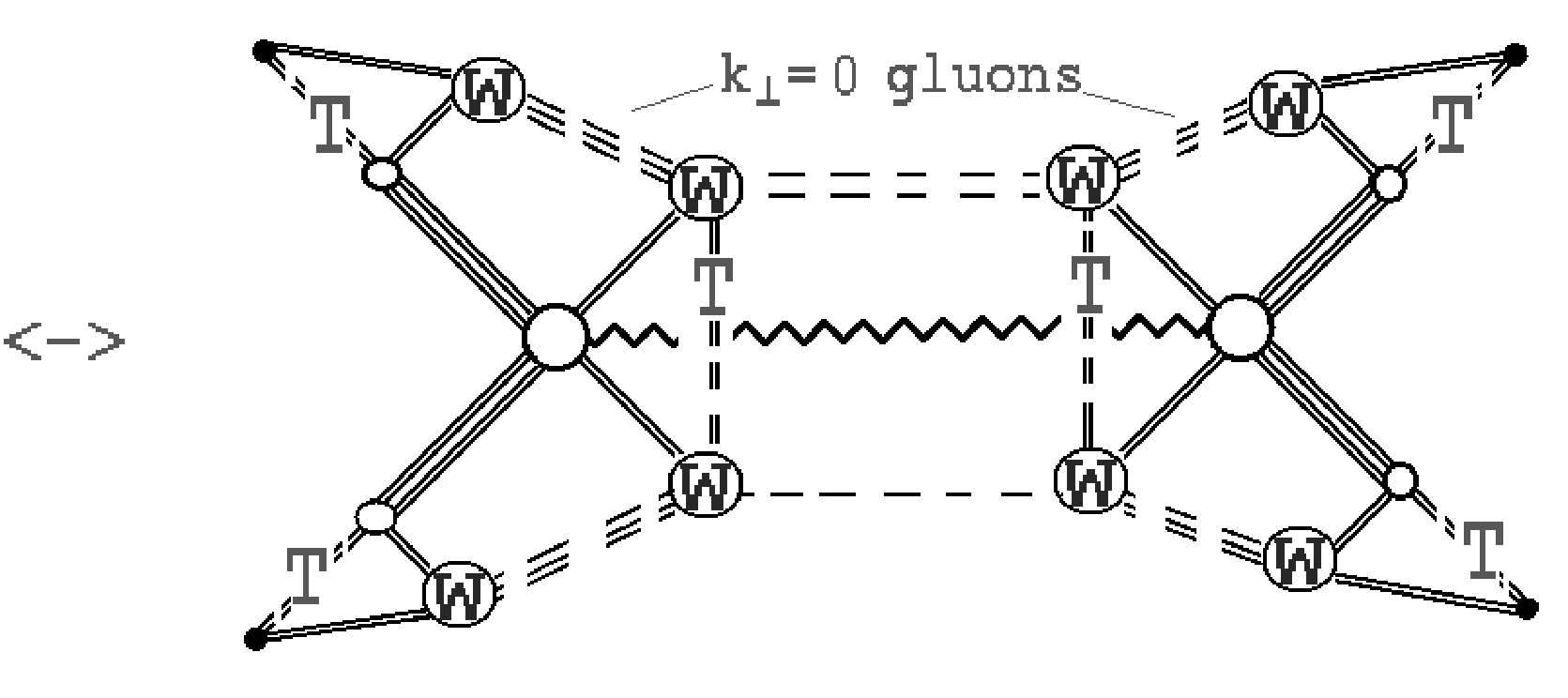}
}
\hspace{0.3in}
\parbox{2in}{
\epsfxsize=1.8in
\epsffile{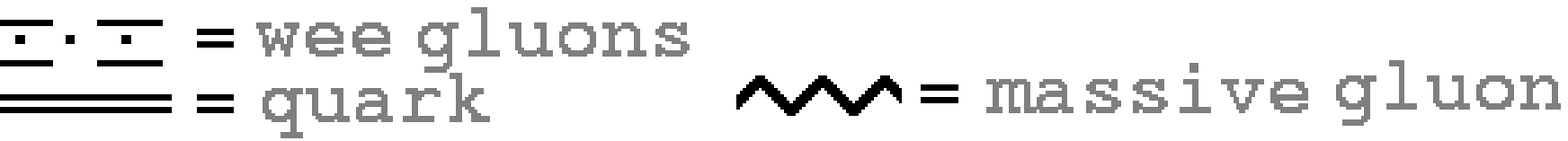}
\epsfxsize=1.9in
\epsffile{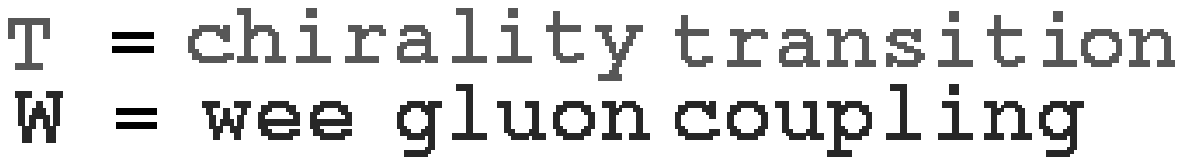}
}
\newline $~$

\centerline{Fig.~1 The Pion Scattering Amplitude}

The pomeron, in  $CSQCD_S$, is 
a reggeized gluon plus the condensate and interactions are described 
by the supercritical pomeron. The asymptotic states are 
\newline \parbox{6in}{\openup-2.5\jot 
\begin{enumerate}
\item{``pions'' $\leftrightarrow$
\{$q\bar{q}$ + wee gluons\}  $\to$ normal meson spectrum in $QCD_S$}
\item{``Pions'' $\leftrightarrow$ 
\{$Q\bar{Q}$ + wee gluons\}  $\to$ $\Pi^{\pm},\Pi^0,$ in  $QCD_S$ }
\item{``nucleons'' $\leftrightarrow$ 
\{$qq~/~\bar{q}\bar{q}$ + wee gluons\} + \{$q~/~\bar{q}$\}, $\to$ SU(3) color singlet 
\newline $~$
\newline $\to$ normal nucleon spectrum in $QCD_S$ }
\item{``Nucleons'' $\leftrightarrow$
\{$QQ~/~\bar{Q}\bar{Q}$ + wee gluons\} + \{$Q~/~\bar{Q}$\}, 
$\to$ $N_6$, $P^{\pm}_6$ in $QCD_S$ }
\end{enumerate}}
$~~~~~~~$ The states and amplitudes of $QCD_S$
are obtained by removing the $k_{\perp}$ cut-off and 
restoring SU(3) gauge symmetry via the critical pomeron
phase transition. As part of the transition, 
the condensate disappears and the 
shifting of the Dirac sea becomes dynamical! Simultaneously, 
the SU(2) singlet gluon becomes massless and decouples.
The only remnant of the symmetry breaking is  
anomaly couplings involving (longitudinal) wee gluons.
The pomeron is (approximately)
a short-distance, gauge-invariant, reggeized gluon combined with 
a color compensating, dynamical, anomalous, wee gluon contribution.
``pions'' and ``Pions'' have the same 
wee gluon component, but with a short-distance quark-antiquark pair.
It is very important that, because of the color antisymmetry of  
$qq$ nucleons, there are no $qq\bar{Q}$ hybrids. Consequently,
either the $N_6$ or the $P_6$ has to be stable.

Shortly, we will discuss hard diffractive interactions of the 
pomeron. Because wee gluon interactions are not involved, we can still
represent the wee gluons as a 
zero transverse momentum ``condensate''. (In reality the wee gluons give
a much more complicated dynamical contribution 
over a range of infra-red transverse momenta.) 

\subhead{5. The Sextet QCD Scale and Electroweak Masses.}

The wee gluons of an infinite momentum pion reproduce vacuum effects. 
Via the anomaly, wee gluon interactions of the form shown in Fig.~2 produce
a $W$ mass
\newline \parbox{3.4in}{ $M^2 ~\sim~ g_W^2 \int k dk $ where
$k$ is a wee gluon momentum. (The left-handed $W$ coupling is crucial.)
 From Feynman
graph color factors, we expect
triplet and sextet quark momentum scales 
to be related (approximately) by ``Casimir Scaling'' so that,   
if the wee gluon coupling ($\int k dk$) 
to triplet and sextet quarks is $F_{\Pi}^2$ and $F_{\pi}^2$,
respectively, 
$$
C_6~\alpha_s (F_{\Pi}^2)~\sim ~C_3 ~\alpha_s(F_{\pi}^2)
~~~~C_6/C_3 ~\approx~ 3
$$ 
}
\parbox{2.5in}{
$~$ \epsfxsize=2.4in 
\epsffile{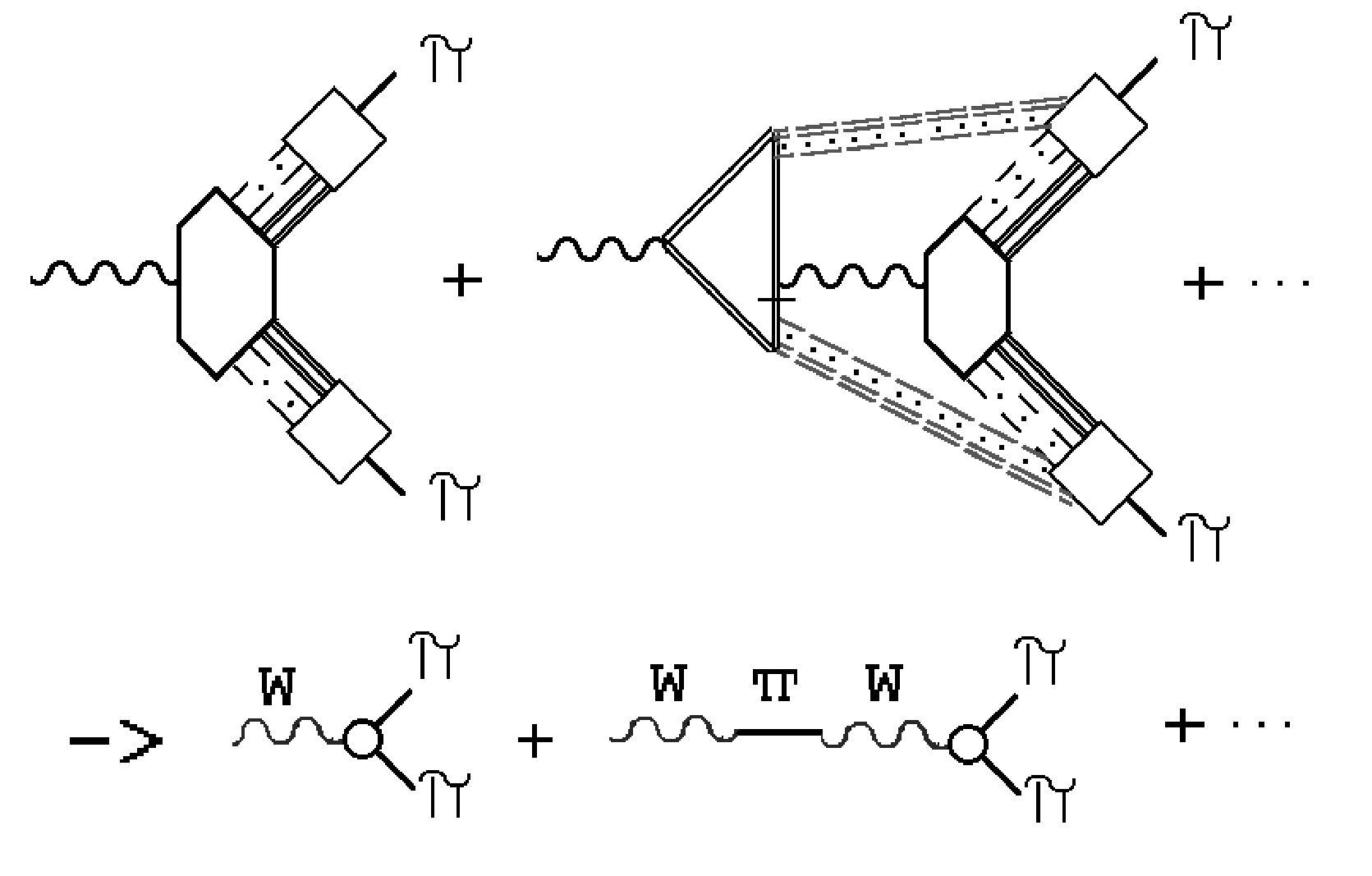}
\centerline{Fig.~2 $~W$ mass generation}}
\newline
Consequently, 
$M$ will be dominated by sextet quark anomaly contributions in Fig.~2
and the
mass generation can be interpreted as due to the $W - \Pi$ coupling,
as illustrated.
If $F_{\pi}$ is the usual triplet chiral scale and 
$\alpha_s$ evolves sufficiently slowly (e.g. $
\alpha_s (F_{\pi}^2)
\sim 0.4~$) $F_{\Pi}$ can indeed be the electroweak scale !
Note that this implies the wee gluon component of the pomeron
couples very strongly ($\sim F_{\Pi}$) to sextet quarks.

\subhead{ 6. Large $x$ and $Q^2$ at HERA.}

An anomaly pole Pion can be produced via a  
large $k_{\perp}$ ``hard interaction'' of the pomeron with a 
$ \gamma$, $Z^0$, or $W^{\pm}$, as illustrated in Fig.~3 for a $\gamma$.
$M_6$ is a dynamical sextet quark mass that, in the following, we will simply 
identify with $F_{\Pi}$. The pomeron provides, directly, 
the wee gluon component that is needed for a massless Pion to appear via 
the anomaly pole. Could the $~\gamma Z^0~ \pom~$ vertex be seen at HERA ?

\noindent \parbox{2.3in}{
\epsfxsize=2.1in
\epsffile{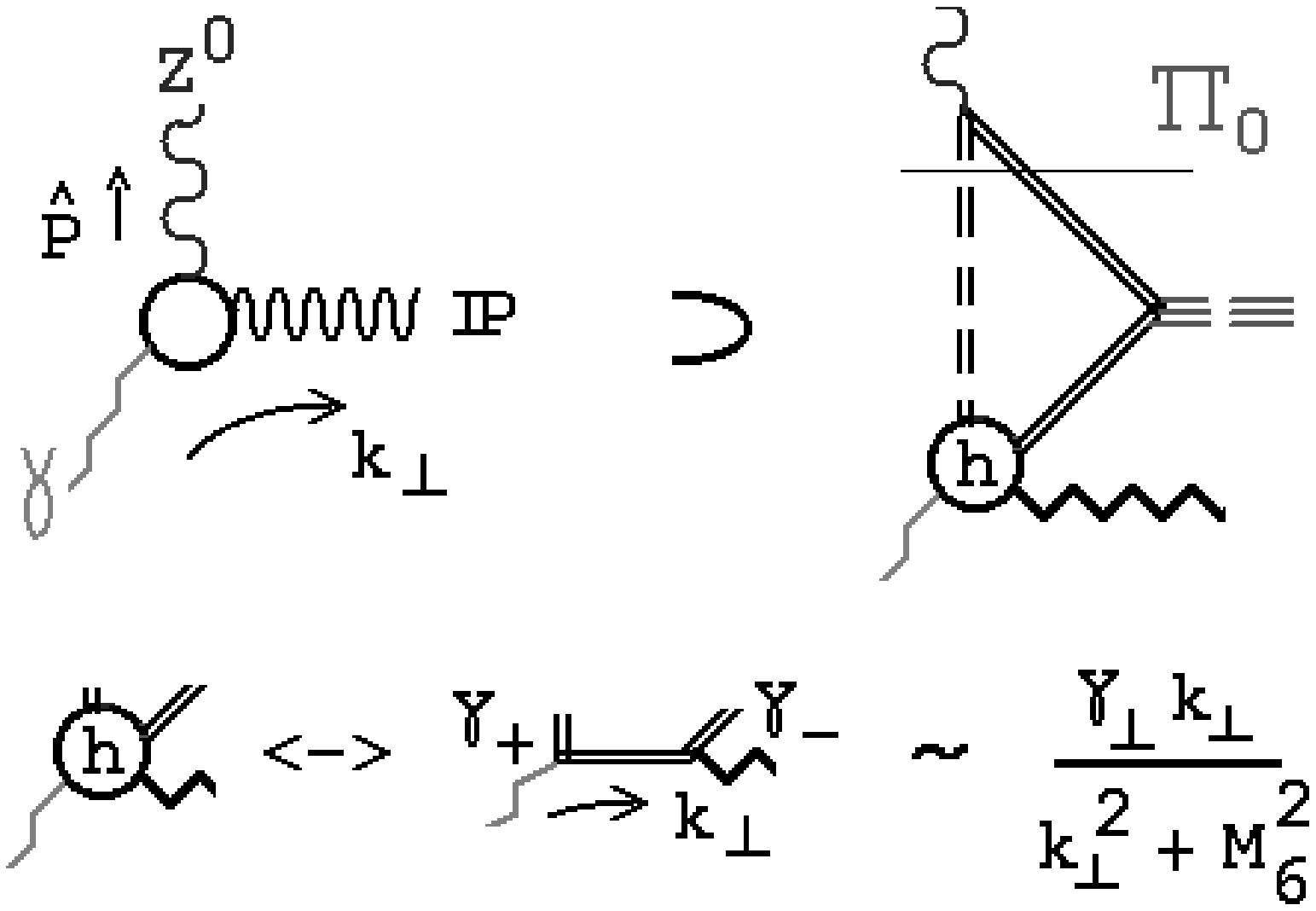}
\begin{center}
Fig.~3 ~The  $\gamma Z^0~ \pom~$ vertex
\end{center}}
\parbox{3.7in}{
Using the anomaly amplitude $
(\hat{P}_+\hat{P}_-\hat{P}_+)/(\hat{P}_-
\hat{P}_+) = \hat{P}+$ to 
extrapolate from $ \hat{P}^2\neq 0$
and combining it with the wee gluon coupling $F_{\Pi}$, 
the $Z^0$ propagator, and vertices $g_w$, gives (using 
$M= g_w F_{\Pi}$)
$$ 
F_{\Pi} 
\hat{P}_+ g_w^2 \frac{(g_{-\nu} - \frac{\hat{P}_-\hat{P}_{\nu}}{M^2})}{
(\hat{P}^2-M^2)}
= - \frac{\hat{P}_-}{F_{\Pi}}\frac{\hat{P}^2~\delta_{-, \nu}}{\hat{P}^2 -M^2}
 - \frac{\hat{P}_+ ~\delta_{+,\nu} }{F_{\Pi}}
$$
The first term (which is present as soon as $ \hat{P}^2 \neq 0$) produces 
physical, longitudinal, $W$'s and $Z$'s. The} 
second term has no pole, but 
provides the background cross-section $\sigma_{\Pi}$, on top of which the 
$Z$ peak appears. Combined with the hard interaction $h$
it gives a factor $(\hat{P}_{+}/F_{\Pi})(\gamma_{\perp}\cdot k_{\perp}
k_{\perp}^2 + M_6^2) \sim \hat{P}_+ k_{\perp}/F_{\Pi}^3$.
(Note that when $ \hat{P}_+ \sim F_{\Pi}$, and $\hat{P}_- << \hat{P}_+$,
this term gives a direct coupling  
to fermion final states proportional to fermion masses.)

If we consider the full DIS amplitude shown in Fig.~4(a) 
and compare it with 
\newline \parbox{2.8in}{the standard two jet amplitude shown in Fig.~4(b),
the only difference, apart from
compensating gap and color factors, is that 
a (triplet) quark 
propagator carrying
momentum $P_j \sim k_{\perp}$ replaces the sextet
triangle diagram factor 
$\sim \hat{P}_+  k_{\perp}/ F_{\Pi}^3~$ (for the 
background amplitude). Therefore,  
$\hat{P}_+  \sim k_{\perp}\sim F_{\Pi}, ~\to ~\sigma_{\Pi} \sim  
\sigma_{2j}$.}
\hspace{0.2in}
\parbox{3in}{
\epsfxsize=1.4in
\epsffile{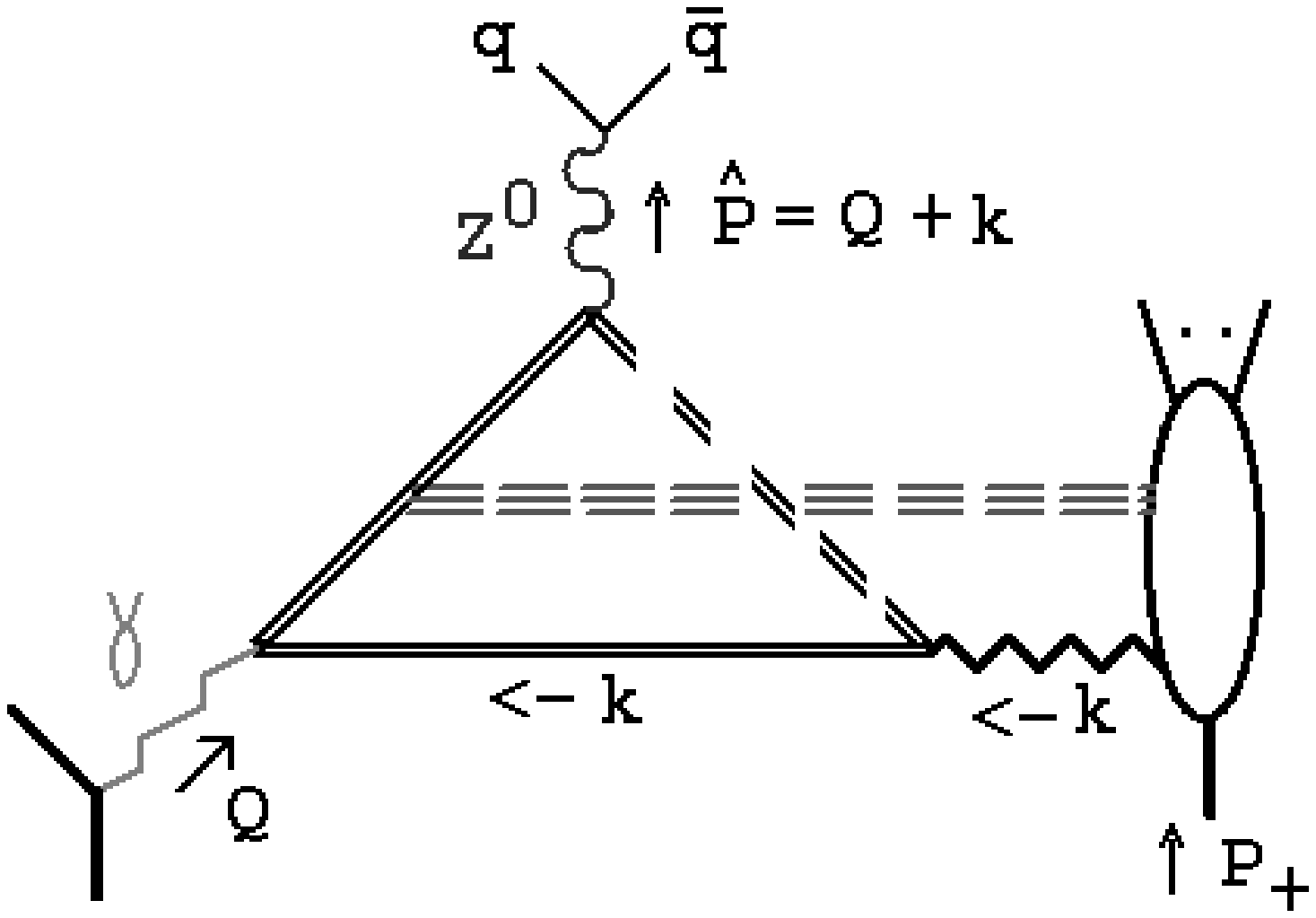}
\hspace{0.1in}
\epsfxsize=1.3in
\epsffile{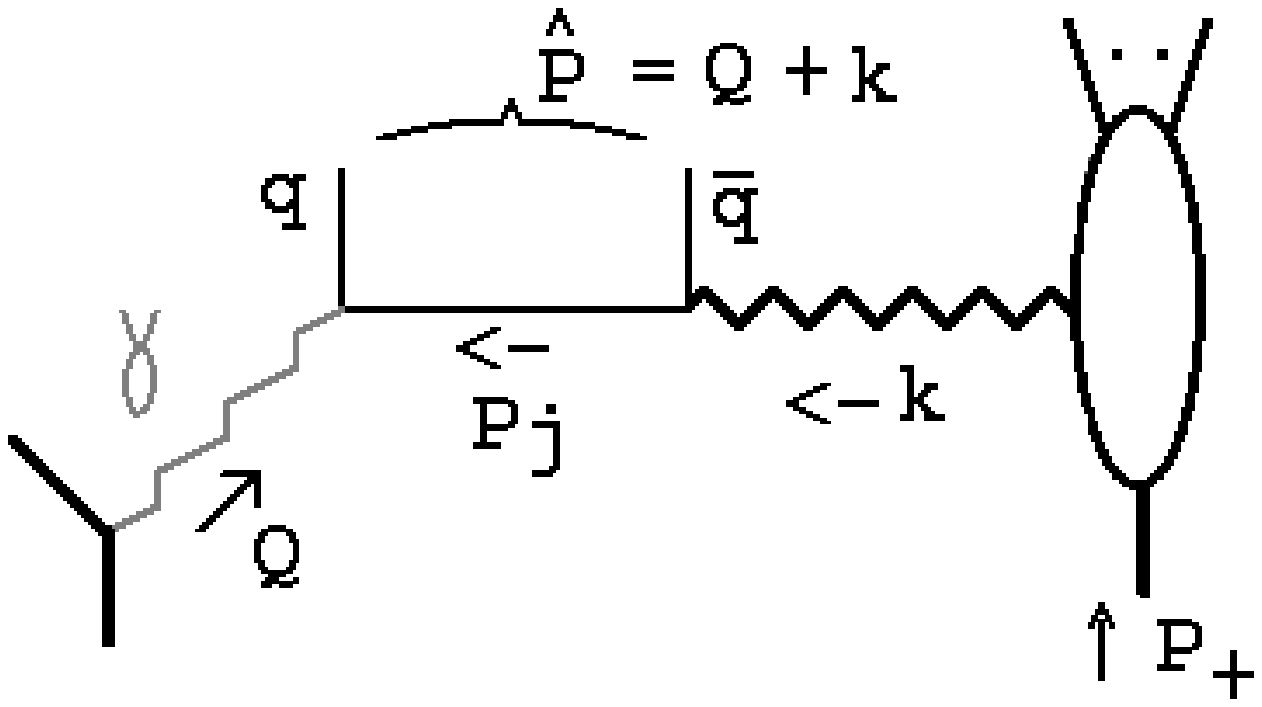}

\centerline{(a)\hspace{1.6in}(b)}
\begin{center}
Fig.~4 (a) $Z^0$ and (b) jet amplitudes 
\end{center}}

 At HERA, large $Q_{\perp}$ requires large  $x$ and  $Q^2$.  
$~ Q_{\perp} ~\centerunder{$>$}{$\sim$}~ 100~ GeV$ requires 
\newline $Q^2 ~\centerunder{$>$}{$\sim$}~ 30,000~GeV^2,~ 
~x~ \centerunder{$>$}{$\sim$}~~ 0.5$.
But $k_{\perp}~ \centerunder{$>$}{$\sim$}~ 100~ GeV~$
and $~\hat{P}^2 ~\sim ~M_{Z^0}^2~ $, requires 
\newline $ |t|~ \centerunder{$>$}{$\sim$}~ ~2k_{\perp}^2~ \sim~  20,000~GeV^2$ 
and $~\sigma_{2j}~$ will be much too small. However, the 
\newline ``non-perturbative'' 
$\gamma Z^0~ \pom~$ vertex should decrease only slowly (with a 
scale determined by $M_6$) as $k_{\perp}$ (and $|t|$) decreases.
Therefore, the increase of the proton/pomeron coupling 
as $|t|$ decreases, combined with 
the contribution of the $Z^0$ pole could give 
an observable jet cross-section.

\subhead{7. The $\eta_6$, $t\bar{t}$, and Large $E_T$ Jets.}

In $CSQCD_S$, the $\eta_6$ has two anomaly couplings to wee gluons.
As illustrated 
\newline \parbox{1.7in}{
\epsfxsize=1.5in
\epsffile{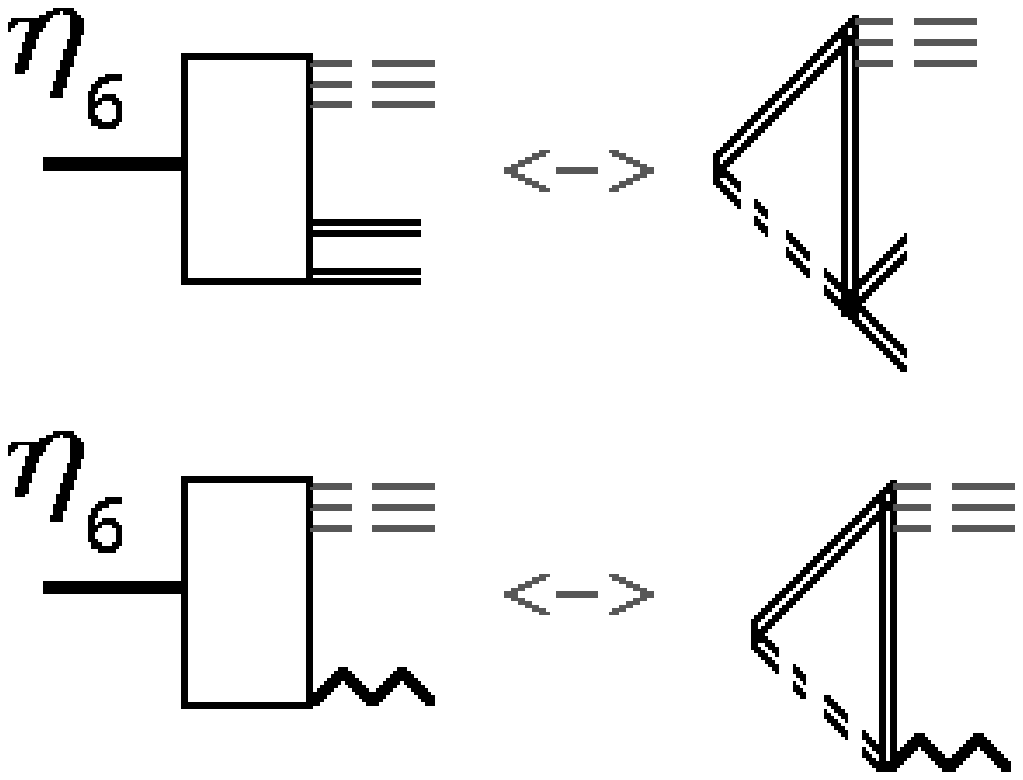}
\centerline{Fig.~5 $\eta_6$ couplings }}
\parbox{4.3in}{in Fig.~5, there is both 
a $Q\bar{Q}$ and an SU(2) singlet gluon coupling (where 
the gluon has a non-leading helicity). 
Therefore, in $QCD_S$, the $\eta_6$ 
mixes with a pure glue state and so should have 
an electroweak scale mass determined, essentially, by $M_6$.
The $\eta_6$ also mixes, via the gluon state, with the triplet flavor singlet 
($\eta_3$) that will be dominated by $t\bar{t}$ at the electroweak scale.
Consequently, the
electroweak scale short-distance component of the $\eta_6$ (which 
carries octet color that is compensated by wee gluons) can be 
produced via gluon production and, since 
sextet quarks are stable, it will decay,} primarily, through 
$t\bar{t}$. Therefore, ``perturbative'' 
$t\bar{t}$ production at Fermilab could be
\newline \parbox{2in}{
\epsfxsize=1.9in
\epsffile{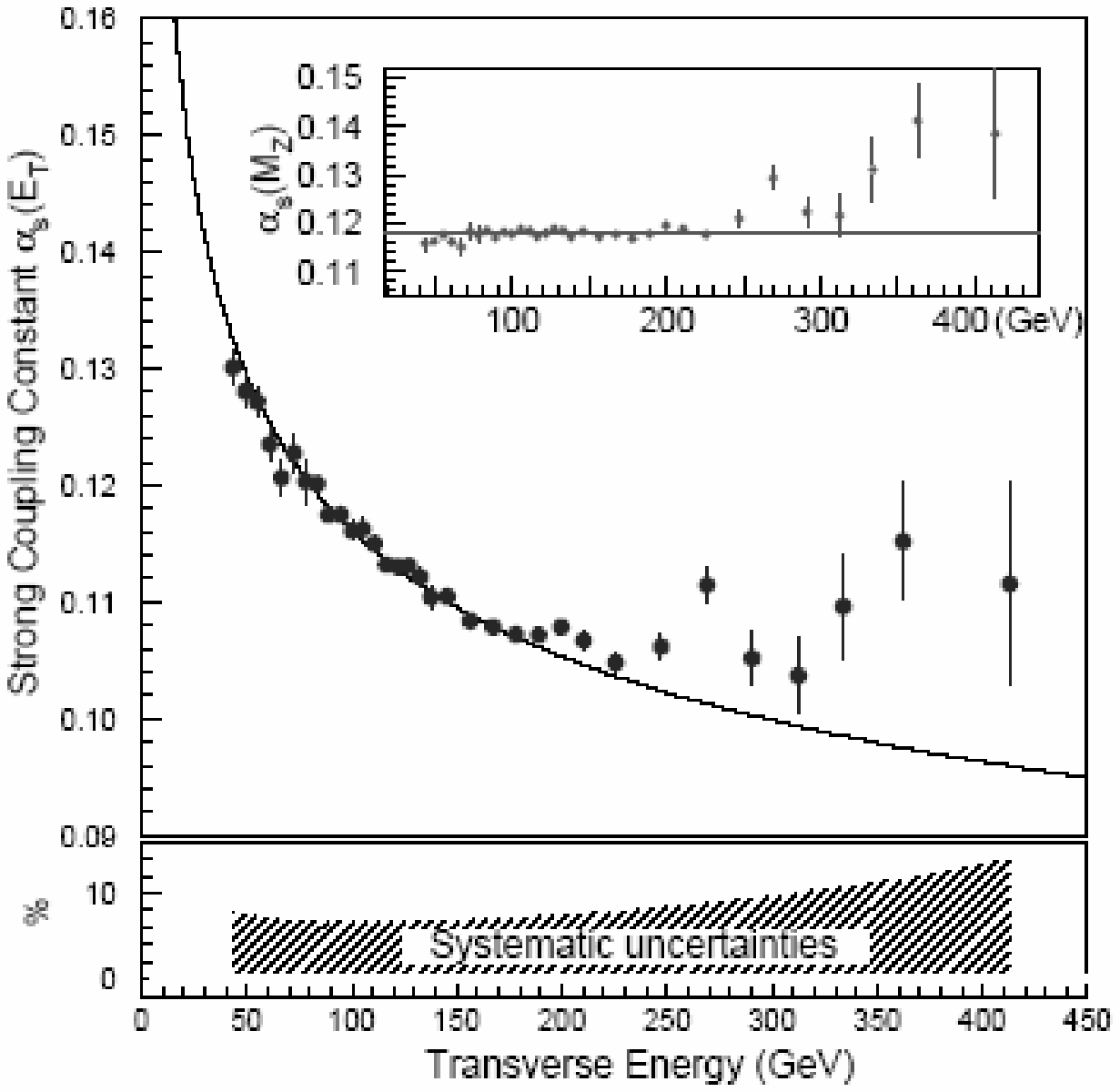}
\begin{center}
Fig.~6 $\alpha_S$ at large $E_T$
\end{center}}
\parbox{4in}{due  to the $\eta_6$, 
with $m_{\eta_6} \sim $ ``$~2 m_t~$''. This would resolve 
the paradox of the apparent production of a confined, colored, quark. 
The sextet doublet ($\equiv$ 10 triplets) 
would then halt the evolution of $\alpha_s$ at $E_T \sim m_t$, implying
a jet excess above $E_T \sim m_t$ of the kind that, as shown in Fig.~6, 
is apparently seen in the data$^2$.
If the $\eta_6$ is indeed responsible for $t\bar{t}$ production, then
we would also expect to see ``non-perturbative'' decay modes of the form   
$$
\eqalign{ \eta_6 &\to W^+W^-Z^0 
\to W^+ W^- b \bar{b}~~ /~ \to W^+ W^- c \bar{c}~~/~ \cdots~,\cr
\eta_6 &\to Z^0Z^0Z^0~,~~\eta_6 \to W^+W^-\gamma~,~~ ...}
$$}

\subhead{8. Diffractive-Related Physics at Fermilab.}

The Tevatron energy is probably too low for the pure pomeron 
production of $W$ and $Z$ pairs that, as we discuss below, we expect to see 
at the LHC.
Single diffractive production of a $W$ or $Z$ can proceed via $WW~\pom~$ 
and $ZZ~\pom~$ vertices that are analagous to the $\gamma Z~\pom~$ vertex
but, because an initial (perturbative) $W$ or $Z$ is required, 
cross-sections will be relatively small. However,
the $t$ dependence of the pomeron/hadron vertex 
implies there should be ``relatively large'' forward
\newline \parbox{4.1in}{
\parbox{4in}{ cross-sections. Also, 
anomalous rapidity
dependence in $W/Z$ + jet cross-sections
is expected.}
\newline $~$
\newline \parbox{4in}{
$~~~~~~~$ Diffractive production of a $W$ or $Z$ pair via a double anomaly 
pole vertex, as illustrated in Fig.~7, 
may give the most clearly anomalous cross-section (although still 
small). Unfortunately, the hadronic decays involved are 
difficult to detect because of the large QCD background 
of $W$ (or $Z$) plus two jets. Related $W$ or $Z$ pair events
}}
\parbox{1.8in}{
\epsfxsize=1.8in
\epsffile{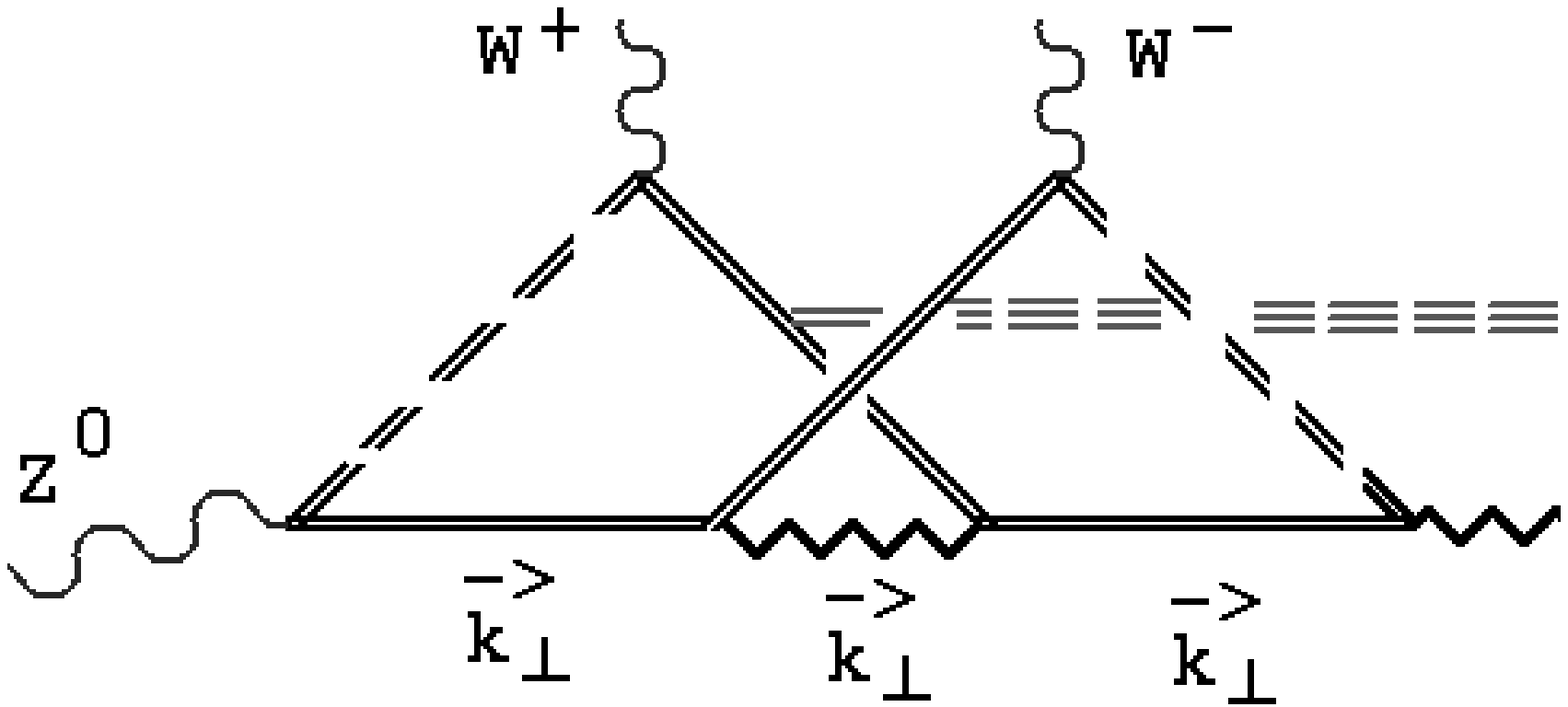}
\begin{center}
Fig.~7 $~Z^0~\pom~ \to W^+W^-$
\end{center}}
\newline with unexpectedly low 
associated multiplicity (anticipating the higher-energy double pomeron
cross-section), or high multiplicity (via AGK), should contribute to 
an inclusive cross-section that, as we discuss next, becomes 
very large at higher energies.
 
\subhead{9. Dramatic Physics at the LHC. }

If the sextet sector exists, the LHC will most probably 
be the discovery machine. The first evidence is likely to
be that, in general, diboson cross-sections
are much larger than expected. However,  
the double pomeron cross-section may well be 
the most definitive early evidence.

\noindent {\bf 9.1 Double Pomeron Exchange.}
Pions can be pair-produced 
directly in double pomeron exchange 
via the anomaly mechanism illustrated in Fig.~8.
An order-of-magnitude argument, analagous to that discussed
above, says that the production of 
\newline \parbox{1.7in}{
\epsfxsize=1.5in
\epsffile{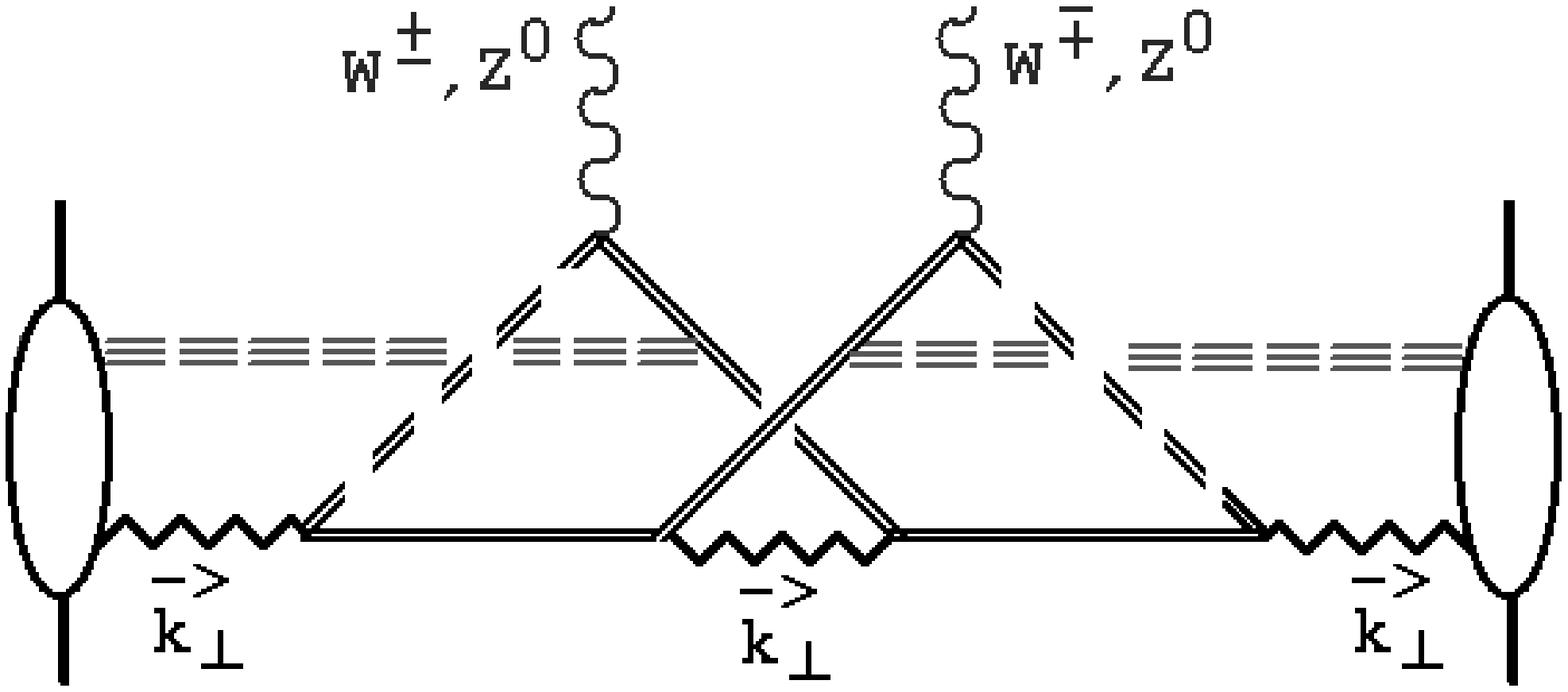}
\begin{center}
Fig.~8 $~\pom~~\pom~ \to \Pi~\Pi$
\end{center}}
\parbox{4.2in}{(large $k_{\perp}$) 
$W^+W^-$ and $Z^0Z^0$ pairs should give jet
cross-sections that are as large as those predicted by standard QCD.
There should also be top quark production via
the ``background'' anomaly vertex and if new leptons exist, with 
electroweak scale masses, there will be
similar vertices for their production. While the central double pomeron vertices }
\newline should vary only slowly with $k_{\perp}$, the external 
hadron/pomeron vertices  
will have strong $ k_{\perp}~$- dependence and give large 
cross-sections at small $t$. 
With forward protons tagged, as planned, it should be immediately seen that 
the double pomeron cross-section for vector boson
pairs is excessively large and, moreover,
has the strong $t$ dependence of a typical hadronic pomeron cross-section.
Possibly, there could be spectacular events 
in which the forward protons are tagged and only
large $E_T$ leptons are seen in the central detector  !!

\noindent {\bf 9.2 Inclusive Cross-Sections for Sextet States.}
If double pomeron couplings are large then 
``cut pomeron'' amplitudes, of the kind illustrated in Fig.~9, that describe the 
central region inclusive production of a
$W^{\pm}$ or $Z^0$ pair will also be large. 
\newline 
\parbox{1.6in}{
\epsfxsize=1.5in
\epsffile{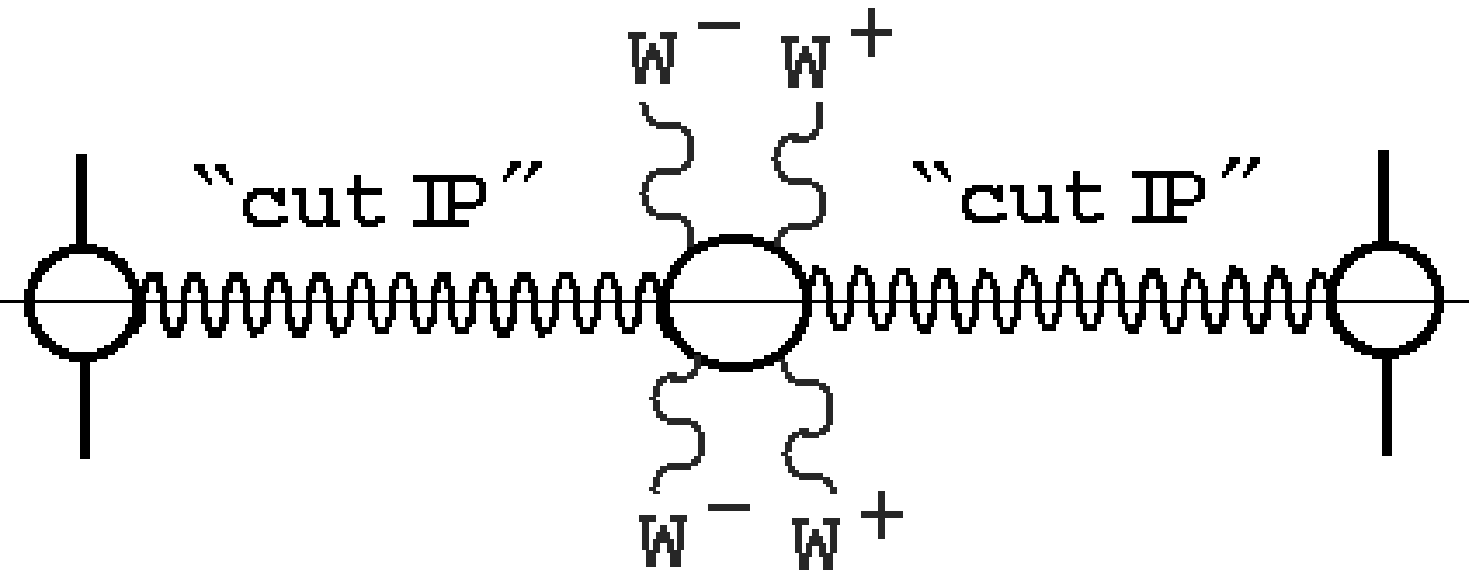}
\centerline{Fig.~9 ~Cut $~\pom~$'s}}
\parbox{4.3in}{Therefore, the initial 
``major change'' in the strong interaction that we expect 
above the electroweak scale is that $W$ and $Z$ (sextet pion)
pairs will be multiply, and strongly, produced across
most of the rapidity axis - in close analogy
with pion production at much lower energies.}

Although $N_6$ and $P^{\pm}_6$ pairs (and also 
$\eta_6$ pairs) may be too massive to be produced in double pomeron
exchange, central region inclusive cross-sections should be large, implying
that ``dark matter'' (see below), in the form of stable, massive, neutral, strongly
interacting particles, should be copiously produced.

\subhead{10. Dark Matter and Cosmic Ray Physics }

Sextet quark current masses must be zero. If not, Pions would 
be massive and could not mix with massless $W/Z$ states to give them masses.
Therefore, the $P_6$ - $N_6$ mass difference is entirely electromagnetic in origin
and (without unification) the $N_6$ is neutral, stable, 
and presumably very massive ($\sim$ 500 GeV, 1TeV ?). It does not form
bound states with normal quark states and
would have been very 
strongly produced in the early universe. It is a natural candidate
to form dark matter (nuclei, clumps, .. ?) 

The well-known ``knee'' in the cosmic ray spectrum, 
shown$^3$ in Fig.~10(a), suggests 
a major interaction change, between Tevatron and LHC energies,
that could be produced by the orders of magnitude 
rise in the $W$ and $Z$ pair production cross-section 
described above. Because the average transverse momentum increases
dramatically, an increasing fraction of produced particles will miss the
detectors. Also an increasing amount of
energy will go into neutrinos (and, at higher energy, $N_6$ pairs) that are not
\newline \parbox{4in}{
\epsfxsize=1.7in 
\epsffile{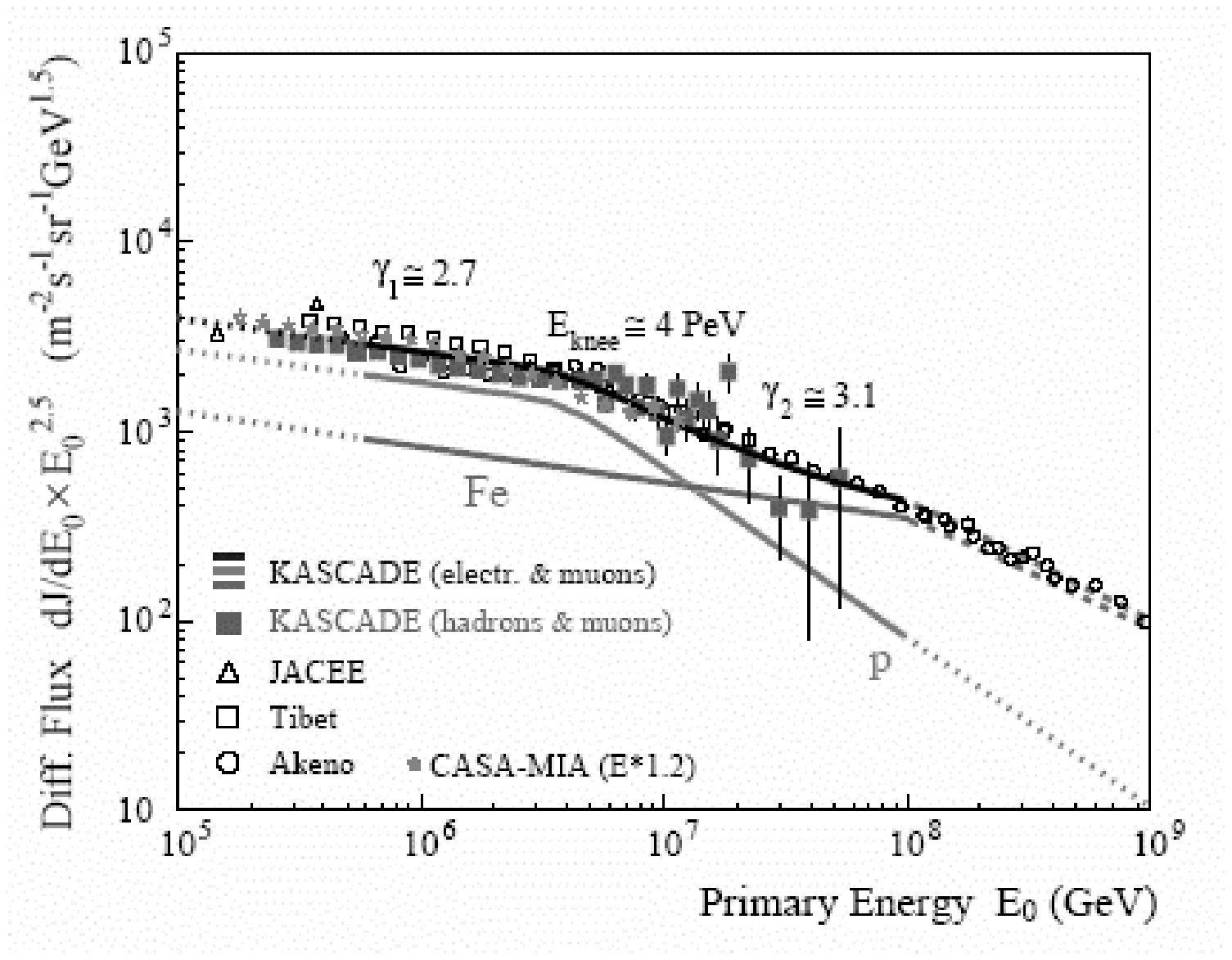}
\epsfxsize=2.1in 
\epsffile{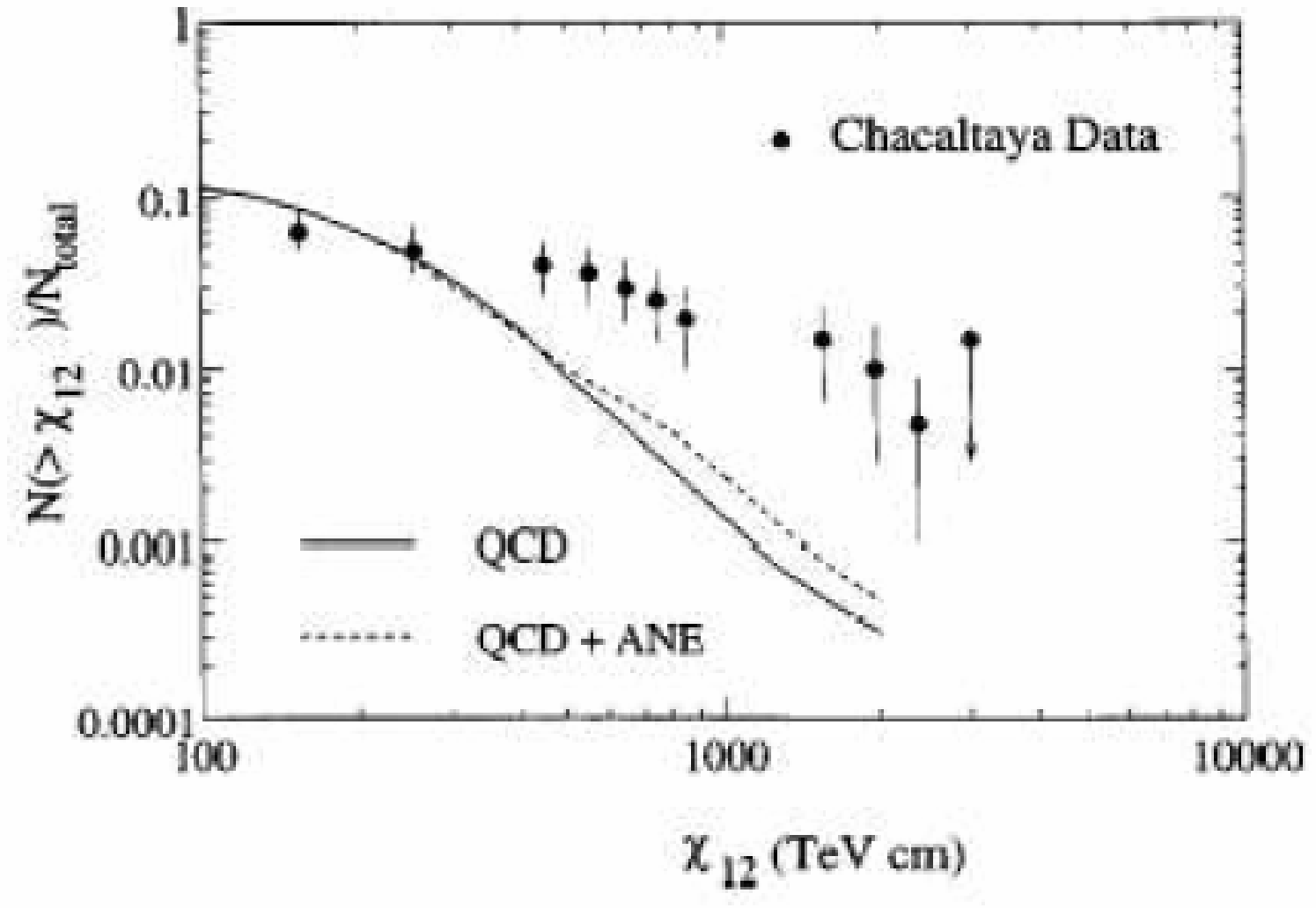}
\begin{center}
(a)\hspace{1.9in}(b)
\newline $~$
\newline Fig.~10 (a) the knee (b) ``dijets''
\end{center}}
\parbox{1.9in}{detected. Serious
underestimation of the total energy will then give an effective ``knee''
in the spectrum. Indeed, as illustrated in Fig.~10(b), 
large transverse momenta ``dijet'' events 
have been seen$^4$ with a cross-section that is 
orders of magnitude larger than in conventional QCD.}
\newline Because neutral, massive, $N_6$'s will 
avoid the GZK cut-off they could also be 
ultra high-energy cosmic rays~! Since they would simply be
very high energy dark matter, their origin would, presumably, 
no longer be a mystery.

\noindent {\bf References} 

\noindent 1. A complete set of
references is given in hep-ph/0405190. A full length paper 
is in preparation, containing a detailed presentation of
all the topics covered in this paper. 
\newline 2. CDF Collaboration (T. Affolder et al.), 
{\it Phys. Rev. Lett.} {\bf 88} 042001 (2002).
\newline 3. M.~Boratav and A.~ Watson, astro-ph/0009469.
\newline 4.  Z.~Cao, L.~K.~Ding, Q.~Q.~Zhu, Y.~D.~He,
{\it Phys. Rev.} {\bf D56} 7361 (1997).

\end{document}